\RequirePackage{etoolbox}
\patchcmd{\bibliographystyle}{#1}{abbrvnat}{}{}
\documentclass[hf]{ceurart}
\usepackage[utf8]{inputenc}
\usepackage[T1]{fontenc}
\usepackage{cwdefs}
\usepackage{caption}
\usepackage{amssymb}
\usepackage{amsfonts}
\usepackage{amsmath}
\usepackage{bm}
\usepackage{xspace}
\usepackage{booktabs}
\usepackage{tikz}
\usetikzlibrary{matrix,snakes,arrows,shapes,positioning,arrows.meta,shapes.geometric}

\hypersetup{breaklinks=true}
\usepackage{fancyvrb}

\newcommand{\nhphantom}[1]{\sbox0{#1}\hspace*{-\the\wd0}}

\mathchardef\hyph="2D
\renewcommand{\P}{\f{P}}
\renewcommand{\c}{\f{c}}

\renewcommand{\c}{\f{i}}

\newcommand{\D}{\f{D}}

\newcommand\subsumedBy{\mathrel{\ooalign{$\geq$\cr
      \hidewidth\raise.225ex\hbox{$\cdot\mkern7.0mu$}\cr}}}
\newcommand\subsumes{\mathrel{\ooalign{$\leq$\cr
      \hidewidth\raise.225ex\hbox{$\cdot\mkern2.0mu$}\cr}}}
\newcommand\strictlySubsumedBy{\mathrel{\ooalign{$\geq$\cr
      \hidewidth\raise.225ex\hbox{$\cdot\mkern7.0mu$}\cr}}}
\newcommand\variant{\mathrel{\ooalign{$=$\cr
      \hidewidth\raise.7ex\hbox{$\cdot\mkern4.5mu$}\cr}}}

\newcommand{\Det}{\text{\textit{Det}}\xspace}

\newcommand{\Lukasiewicz}{{\L}u\-ka\-sie\-wicz\xspace}

\newcommand{\ProverN}{\textit{Prover9}\xspace}
\newcommand{\OTTER}{\textit{OTTER}\xspace}
\newcommand{\E}{\textit{E}\xspace}

\newcommand{\CMProver}{\textit{CMProver}\xspace}
\newcommand{\SETHEO}{\textit{SETHEO}\xspace}
\newcommand{\PIE}{\textit{PIE}\xspace}
\newcommand{\CDTOOLS}{\textit{CD~Tools}\xspace}
\newcommand{\SGCD}{\textit{SGCD}\xspace}
\newcommand{\CCS}{\textit{CCS}\xspace}

\newcommand{\TPTP}{\textit{TPTP}\xspace}
\newcommand{\TPTPCDT}{\mbox{\textit{TPTPCDT2}}\xspace}

\newcommand{\tp}[1]{\textsf{#1}\xspace}

\newcommand{\trskip}[1]{\ldots}

\newcommand{\ddim}[3]{\la #1, #2, #3\ra}

\renewcommand{\fimp}{\Rightarrow}

\newcommand{\mereq}{\hspace{0.15em}=\hspace{0.15em}}
\renewcommand{\n}{\mathrm{n}}

\newcommand{\code}[1]{\texttt{#1}}

\renewcommand{\P}{\f{thm}}
\renewcommand{\P}{\f{Thm}}

\makeatletter
\RenewDocumentCommand \maketitle {} {
  \thispagestyle{plain}
  \xdef\firstpage{\thepage}
  \ifbool{longmktitle}
  {
    \LongMaketitleBox
    \ProcessLongTitleBox
  }
  {
    \ifbool{dc}
    { \twocolumn[\MaketitleBox] }
    { \MaketitleBox }
    \printFirstPageNotes
  }
  \normalcolor \normalfont
  \renewcommand\thefootnote{\arabic{footnote}}
  \gdef\@pdfauthor{\infoauthors}
  \hypersetup{%
    pdfcreator={ceurart.cls},
    linkcolor={hscolor},
    urlcolor={hscolor},
    citecolor={hscolor},
    filecolor={hscolor},
    menucolor={hscolor},
  }    
}
\makeatother

\renewcommand{\rimp}{\rightarrow_{\mathrm{rew}}}
\renewcommand{\c}[1]{\bm{\mathsf{#1}}}
\begin{document}

\title{\name{CD Tools} -- Condensed Detachment and
  Structure Generating Theorem Proving (System Description)}

\author{Christoph Wernhard}[%
  email=info@christophwernhard.com]
\address{University of Potsdam, Germany}

\begin{abstract}[]
  \CDTOOLS is a Prolog library for experimenting with condensed detachment in
  first-order ATP, which puts a recent formal view centered around proof
  structures into practice. From the viewpoint of first-order ATP, condensed
  detachment offers a setting that is relatively simple but with essential
  features and serious applications, making it attractive as a basis for
  developing and evaluating novel techniques. \CDTOOLS includes specialized
  provers based on the enumeration of proof structures. We focus here on one
  of these, \SGCD, which permits to blend goal- and axiom-driven proof search
  in particularly flexible ways. In purely goal-driven configurations it acts
  similarly to a prover of the clausal tableaux or connection method family.
  In blended configurations its performance is much stronger, close to
  state-of-the-art provers, while emitting relatively short proofs.
  Experiments show characteristics and application possibilities of the
  structure generating approach realized by that prover. For a historic
  problem often studied in ATP it produced a new proof that is much shorter
  than any known one.
\end{abstract}


\maketitle

\section{Introduction}
\label{sec-intro}

We present \CDTOOLS, a Prolog library for experimenting with condensed
detachment (CD) \cite{ulrich:legacy:2001} in automated theorem proving (ATP).
CD is basically a framework for first-order reasoning about propositional
logics, introduced by Carew A. Meredith
\cite{prior:logicians:1956}.\footnote{General first-order proving for Horn
  problems is possible with very similar techniques. For examples, see
  \cite{cw:ccs} and
  \url{http://cs.christophwernhard.com/cdtools/exp-ccs-2022-06/table_4.html}.}
Its inference rule, intuitively modus ponens with unification, may be
described as positive hyperresolution with the single non-unit clause
\begin{equation}
  \tag{\Det}
\P(y) \revimp \P(x\fimp y) \land \P(x).
\end{equation}
The propositional formulas to reason about are represented there by terms,
with $\fimp$ as function symbol for implication, wrapped in the unary
predicate~$\P$.\footnote{Our presentation of the clause (\Det) is oriented at
  the proof structures considered with CD, discussed below. In a more
  conventional ATP-oriented form, with the function symbol for implication in
  prefix notation, (\Det) might be written as $\lnot \P(x) \lor \lnot
  \P(\f{implies}(x,y)) \lor \P(y)$.} In an application of (\Det), $\P(y)$ is
called \defname{conclusion}, $\P(x\fimp y)$ \defname{major premise} and
$\P(x)$ \defname{minor premise}.
A CD proof is a full binary tree or term (\name{term} now understood not as
formula constituent but rather on the meta or proof-structure level) such that
for inner nodes $\D(d_1,d_2)$ the arguments $d_1$ and $d_2$ are the subproofs
of the major and minor premise, and leaves or constants are labels of proper
axioms of the considered propositional object logics, unit clauses, e.g.,
$\P(p\fimp(q\fimp p))$, where $p,q$ are variables. We call these proof
structure representations \defname{D-terms}. See \cite{cwwb:lukas:2021} for a
precise account.

CD is attractive as a basis for research on first-order ATP because it allows
to express application problems, including hard proving problems and questions
for deriving theorems that render mathematical intuition, while requiring only
a simplified setting of first-order ATP, with a single predicate, a single
non-unit clause which is Horn, and no equality. Actually, research with CD
belongs to the first and most successful applications of ATP
\cite{mccune:wos:cd:1992} and numerous techniques for \OTTER \cite{otter}
emerged from it
\cite{wos:contributes:1990,wos:bledsoe:91,mccune:wos:cd:1992,wos:resonance:95,wos:combining:96,veroff:shortest:2001,fitelson:missing:2001,wos:meredith}.
A dedicated system for CD is described in \cite{fuchs:code:1997}. \ProverN
\cite{prover9} appears to handle inputs that represent a CD problem in a
special way by selecting hyperresolution as inference rule such that proofs
may be viewed as CD proofs.
Recently \cite{cwwb:lukas:2021} it was observed that the connection method
\cite{bibel:atp:1982,bibel:otten:2020} suggests an ATP-oriented view on CD
that is not centered on associating with (\Det) an inference rule but on the
proof structure, the \mbox{D-term}, as a whole. A D-term represents the
structure formed by the connections. A proof consists of a D-term together
with a substitution on the terms of conclusion formulas $\P(y)$ associated
with each node. Informally, this substitution is the most general one that is
constrained by copies of axioms at the leaves and copies of (\Det) at the
inner nodes.

\CDTOOLS is motivated by exploring that view. It provides functionality to
inspect, relate and simplify given proofs in various ways and aims to be a
practical basis for developing new first-order ATP techniques. It currently
included two experimental provers, \CCS (\name{Compressed Combinatory Proof
  Structures}) \cite{cw:ccs} and \SGCD (\name{Structure Generation for
  Condensed Detachment}). \SGCD is specialized for CD problems and can (in the
union of four different configurations) solve 93\% of the CD problems in the
\TPTP \cite{tptp} that are solvable by any prover at all.\footnote{According
  to the latest rating values in \name{TPTP~8.0.0}. Considered were the 196 CD
  problems of corpus \TPTPCDT (see Sect.~\ref{sec-overview}).} Its paradigm may
be called \name{structure generating}, because it basically operates, like
goal-driven provers describable in terms of the connection method, model
elimination or clausal tableaux
\cite{pttp,setheo:92,letz:habil,leancop,cw:pie:2016}, by enumerating proof
structures, which here means D-terms. A caching or lemma mechanism permits to
configure the prover in the space between the extremes of purely goal- and
axiom-driven proof search.

The paper is organized as follows: In Sect.~\ref{sec-overview} we overview the
system and specify the considered problem corpus, in Sect.~\ref{sec-sgcd} the
included prover \SGCD is described, and in Sect.~\ref{sec-exp} seven
subsections present experiments that indicate characteristics and application
possibilities of \SGCD and its approach. Section~\ref{sec-conclusion}
concludes the paper. The second included prover, \CCS, mentioned here on
occasion, is described in \cite{cw:ccs}. The system is available as free
software from
\begin{center}
  \url{http://cs.christophwernhard.com/cdtools/}.
\end{center}
That website also provides detailed result tables, full output logs, including
Prolog-readable proof terms, as well as reproduction instructions for the
experiments described in the paper. For some experiments, also graph
visualizations of proofs are shown there.

\section{Overview on the System}
\label{sec-overview}

\CDTOOLS is implemented in \name{SWI-Prolog} \cite{swiprolog}, a free well
maintained modern Prolog system and uses \PIE \cite{cw:pie:2016,cw:pie:2020}
as basic support for first-order ATP.\footnote{No efforts were made to support
  other Prolog systems. Brief tries with free systems reckoned as particularly
  fast, notably \name{YAP} and \name{Eclipse}, unfortunately gave the
  impression that these are broken in their recent versions.} It comprises
about 260 user predicates, modularized in a fine-grained way to keep tight
track of dependencies, into 56~modules, which in turn are grouped into 14 sets
of modules with related or similar functionality.\footnote{See
  \url{http://cs.christophwernhard.com/cdtools/overview.html}.} Prolog is
based on unification, Horn clauses, and realizing the execution of a
nondeterministic program as an enumeration of proofs through backtracking.
These features are quite close to CD and the approach to theorem proving with
enumerating proof structures, which is utilized in \CDTOOLS by building on the
Prolog-provided unification and means for structure enumeration.

The system provides interfaces to several external worlds: It supports
conversion to and from Jan \Lukasiewicz's Polish notation typically used in
the literature on applications of CD, where the axiom $\P(p\fimp(q\fimp p))$,
in Prolog notation \code{thm(P=>(Q=>P))}, is represented by $\g{CpCqp}$.
Predicates are provided to test \TPTP problems whether they represent a CD
problem, and to canonicalize their vocabulary. \name{TPTP~8.0.0} (and
\name{TPTP~7.5.0}\footnote{\name{TPTP~8.0.0} and \name{TPTP~7.5.0} contain the
  same CD problems, with the same difficulty ratings.}) contains 206~CD
problems, all in the \tp{LCL} domain. \CDTOOLS includes a definition of a
slightly smaller problem corpus called \TPTPCDT, comprising the 196 CD
problems in \name{TPTP~8.0.0} that remain after excluding those two with
status \name{satisfiable}, those five with a form of detachment that is based
on implication represented by disjunction and negation, and those three with a
non-atomic goal theorem. The restriction on detachment and goals simplifies
the problem form.

Operations on D-terms in \CDTOOLS are based on the formal treatment and
concepts from \cite{cwwb:lukas:2021}. An important basic way to associate an
atomic formula with a D-term is its \defname{most general theorem (MGT)} (aka
\name{principal type scheme}
\cite{hindley:meredith:cd:1990,hindley:book:1997}) with respect to given
axioms. It is the most general (with respect to subsumption) atomic formula
that is proven with the D-term for the axioms. Figure~\ref{fig-mgt} shows its
implementation in \CDTOOLS.
\begin{figure}[h]
  \caption{Implementation of MGT computation in \CDTOOLS.}
  \centering
\label{fig-mgt}
\begin{Verbatim}
d_mgt(d(DXY,DX), FY) :-
        !,
        d_mgt(DXY, (FX=>FY)),
        d_mgt(DX, FX).
d_mgt(D, F) :-
        id_axiom(D, F),
        acyclic_term(F).
\end{Verbatim}
\end{figure}
The occurs check is realized by the SWI-Prolog library predicate
\code{acyclic\_term/1}. The given axioms are referenced via the predicate
\code{id\_axiom/2}. If this is defined as \code{id\_axiom(1, P=>(Q=>P))}, then
the query \code{?- d\_mgt(d(1,1), X)} succeeds with \code{X} bound to
\code{P=>(Q=>(R=>Q))}, representing the atomic formula
$\P(p\fimp(q\fimp(r\fimp q)))$, where $p,q,r$ are variables. Another way to
use the $\code{d\_mgt/2}$ predicate is to verify that the MGT subsumes a proof
goal given as ground term, as for example in \code{?- d\_mgt(d(1,1),
  p=>((q=>r)=>(p=>(q=>r))))}. The \defname{in-place theorem (IPT)}, where also
substitution constraints induced by a context D-term are taken into account is
a second important way to associate an atomic formula with a D-term, which is
also supported in \CDTOOLS.

The predicate $\code{id\_axiom/2}$ used in the implementation of
\code{d\_mgt/2} represents a global mapping from axiom identifiers to axioms
that is quietly assumed at experimenting, e.g., when computing the MGT of a
given D-term. It is one of the few dynamic (state dependent) predicates of
\CDTOOLS and can be \mbox{[re-]}initialized with an interface predicate. A
pre-defined association of about 170~common names for about 120~well-known
formulas, e.g., from \cite{prior:formal:logic:1962,ulrich:legacy:2001}, helps
to specify these as values of $\code{id\_axiom/2}$ and to detect them among
computed lemmas.

\CDTOOLS includes implementations of many concepts and operations introduced
or discussed in \cite{cwwb:lukas:2021}, including D-term comparison by
$\geq_{\mathrm{c}}$, various notions of regularity, the \name{organic}
property (based on calling a SAT solver\footnote{Via \PIE, which provides a
  call interface to SAT and QBF solvers by translating between \PIE's Prolog
  term representation of formulas and DIMACS or QDIMACS files, commonly used
  by the solvers. In the experiments \name{MiniSat} \cite{minisat:03} was
  used.} or based on previously proven lemmas), the \name{prime} property, and
\name{n-simplification}, that is, replacing complex subproofs of minor
premises whose goal is irrelevant for the conclusion with a primitive
subproof.

D-terms allow precise proof size measures that are supported by \CDTOOLS. We
represent here the \defname{dimensions} of a D-term by a triple
\[\ddim{c}{t}{h}\] that gathers three of its measures: Its \defname{compacted
  size}~$c$, i.e., the number of inner nodes of the unique minimal DAG
representing the tree, its \defname{tree size}~$t$, i.e., the number of inner
nodes (occurrences of the $\f{D}$ symbol, copies of (\Det) used in the proof),
and the \defname{height}~$h$ of the tree. For a multiset of D-terms, $c$ is
the number of inner nodes of the minimal DAG representing the set of its
members, and $t$ is the sum and $h$ the maximum of the respective values of
its members.

\CDTOOLS provides various predicates that enumerate all D-terms of or up to
some given size measure, together with the respective MGT, where D-terms which
have no MGT because the constraints on the substitution are not satisfiable
are omitted. For small values, these predicates can be used to obtain
information about all proofs of a given size. They also can serve as simple
theorem provers, forming the basis of the approach discussed in the next
section and developed further in \cite{cw:ccs}.

\enlargethispage{8pt}
\section{The \SGCD Prover}
\label{sec-sgcd}

\CDTOOLS includes \SGCD, a highly configurable structure generating prover for
\name{CD} problems. The prover is parameterized by a \name{generator}, a
predicate that enumerates all D-terms paired with MGTs at a given
\name{level}. As an example of such a level, consider the tree size of the
D-term. The generator can be used in two modes: \name{goal-driven}, where the
MGT argument is considered as input and instantiated with a ground term, the
goal to prove, and \name{axiom-driven}, where the MGT argument is a variable
that is successively bound to derived lemmas.

In goal-driven mode, the system acts similar to ``\SETHEO-like'' provers
\cite{pttp,setheo:92,letz:habil,leancop,cw:pie:2016}. Invoking the generator
at increasing levels realizes iterative deepening, the usual way of operation
of such systems.
In axiom-driven mode, the generator would be invoked at increasing levels,
where found solutions are cached. Later invocations at higher levels, which
involve enumeration of solutions at lower levels as subproblems, then access
these from the cache instead of recomputing them.

\SGCD combines both modes in the following way: Basically it operates in the
axiom-driven mode, but before a new level is computed and cached, it invokes
the goal-driven mode, at the new level and, depending on the configuration,
possibly at a number of increasingly higher levels. The goal-driven mode has
there access to the cache, which it uses for subproblems that involve finding
solutions below the new level.

For experimenting, various generators have been implemented, for tree size,
height, and other measures, also in variations concerning the order in which
structures appear. Each structure appears only once, although it is of course
possible that different structures prove the same lemma (i.e., have the same
MGT with respect to the given axioms).\footnote{That each structure appears
  only once is not an essential precondition for \SGCD. The level
  characterizations considered so far in experiments could all be
  easily implemented such that each structure is returned only
  once.} The prover can stop after a proof of a given goal has been found or
enumerate alternate proofs. Extreme configurations realize pure axiom-driven
lemma computation and pure goal-driven proof search.
The cached solutions are triples of MGT (a lemma), D-term (its proof) and the
level at which the proof was found. After the axiom-driven mode has completed
a level, the cache is updated with the solutions newly found at the level,
which is configurable in many ways: New solutions whose MGT is subsumed by
some other solution can be deleted. The number of cached solutions can be
given a limit, where the solutions to delete in order to stay within the limit
are determined by an ordering of the union of the old cache and the new
solutions, e.g., according to some size measure of the MGT. If a solution is
deleted from a cached level, it does no longer appear when that level is
accessed for a subproblem of a problem at a higher level. However, deleted
solutions can be kept in an extra store, which can be useful for experimenting
and for applications with axiom-driven theorem generation.

\section{Experiments}
\label{sec-exp}

\subsection{\SGCD and the Condensed Detachment Problems in the \TPTP}
\label{sec-exp-sgcd}

Table~\ref{tab-solnum} summarizes the performance of \SGCD on the \TPTPCDT
problems in comparison to state-of-the-art provers as represented by the \TPTP
rating value:\footnote{\TPTP rating values considered in this paper refer to
  the latest values in \name{TPTP 8.0.0} or in \name{TPTP 7.5.0}, which both
  provides the same values for the considered problems.} A value of 1.00
indicates that the problem is most difficult, i.e., can not be solved by any
state-of-the-art prover (assuming some reasonable resource constraints). Also
own experiments with \name{E~2.6} \cite{eprover} and \ProverN \cite{prover9}.
The latter plays a special role, because for CD problems it chooses by default
hyperresolution as inference rule and thus emits, like \SGCD, actual CD
proofs. Moreover, its output format permits easy translation to D-terms, which
is implemented in \CDTOOLS. As indicated in \cite{veroff:cd:2011}, CD proofs,
in contrast to proofs of other calculi, may be explicitly desirable in
applications.

\begin{table}[h]
  \centering
  \caption{The performance of \SGCD in configurations that
    blend goal- and axiom-driven search.}
  \vspace{4pt}
  \label{tab-solnum}             
  \begin{tabular}{l@{\hspace{1.0em}}r}
    \name{Corpus \TPTPCDT} & 196\\
    \name{Corpus \TPTPCDT}, Rating < 1.00 & 189\\
    \name{Corpus \TPTPCDT}, Rating > 0.00 & 45\\
    \name{Corpus \TPTPCDT}, Rating > 0.25 & 20\\
    \name{E~2.6} & 185\\\midrule
    \name{SGCD-1} & 165\\
    \name{SGCD-1+2+3+7} & 176\\
    \name{SGCD-1+2+3+7}, Rating $>$ 0.00 & 25\\
    \name{SGCD-1+2+3+7}, Rating $>$ 0.25 & 5\\\midrule
    \ProverN & 168\\
    \name{SGCD-1+2+3+7} $\setminus$ \ProverN & 13\\
    \ProverN $\setminus$ \name{SGCD-1+2+3+7} & 5\\
  \end{tabular}
  \vspace{-10pt}
\end{table}

Each row in the table shows for a set of problems its cardinality. The first
five rows describe the corpus and the performance of state-of-the art provers
on it: \name{Corpus \TPTPCDT} is the whole corpus, the subsequent three rows
show its cardinality under restrictions of the problem rating. Row
\name{E~2.6} shows the number of problems in the corpus solved by \name{E~2.6}
\cite{eprover}.\footnote{On a HPC system with
  Intel\textregistered\ Xeon\textregistered\ Platinum 9242 @ 2.30GHz CPUs,
  3.7~GB memory per CPU and 600~s time limit per problem. Option settings were
  \texttt{--auto-schedule --cpu-limit=600}. A time limit of 2400~s with
  --cpu-limit=2400 lead to longer proving times but not to more solved
  problems.}

\SGCD was tested in about two dozens configurations of which
\mbox{\name{SGCD-1}} could solve the most problems. There the generator is
over tree size, goal-driven search is in two levels, solutions where the
dimensions of the MGT exceeds 5~times the maxima in the input are deleted,
solutions with MGTs subsumed by previously found solutions are deleted, the
cache is ordered according to height and size of the MGTs and its size is
limited to 1000 solutions. Row~\mbox{\name{SGCD-1}} shows the number of
problems solved by \SGCD in this
configuration.\footnote{All \label{foot-exp-cond} results for \SGCD reported
  in this subsection were obtained with \name{TPTP 8.0.0} on a HPC system with
  Intel\textregistered\ Xeon\textregistered\ Platinum 9242 @ 2.30GHz CPUs,
  3.7~GB memory per CPU and 2400~s time limit per prover run and problem. If
  results for different configurations are considered together, each of these
  was given the 2400~s time limit.}

\enlargethispage{10pt}

In general, the performance of a first-order prover on a given problem depends
strongly on parameter settings, such that mature systems are typically invoked
in a parallel or portfolio mode with a scheduled pattern of alternate
settings. Since \SGCD does currently not support such a parallel or
portfolio-like invocation, it seems adequate to consider its results for
different configurations grouped together. \mbox{\name{SGCD-2}},
\mbox{\name{SGCD-3}} and \mbox{\name{SGCD-7}} supplement configuration
\mbox{\name{SGCD-1}} with slightly varied settings, in particular a generator
over height, a larger cache size of~3000, and differently characterized
limitations of the MGT dimensions. Row \name{SGCD-1+2+3+7} shows the number of
problems solved in at least one of these four configurations, each tried with
a timeout of 2400~s. Further rows show the number of the solved problems whose
rating is above specific thresholds. None of them has rating 1.00.
Row \ProverN shows the number of problems solved by \ProverN in default
configuration.\footnote{Performed on the same hardware as those with \SGCD,
  also with a timeout of 2400~s for each problem, supplemented for problems
  \tp{LCL020-1} and \tp{LCL021-1} with results obtained on a notebook with 16
  GB RAM and Intel\textregistered\ Core\texttrademark\ i7-8550U @ 1.80GHz CPU,
  because these caused memory exhaustion on the HPC system.} Two further rows
indicate the numbers of problems that can be solved either by \SGCD in at
least one of the four considered configurations or by \ProverN, but not both.
The set of problems solved by \name{E~2.6} includes those solved by \SGCD as
well as those solved by \ProverN.

In summary, the experiment show that from the \TPTPCDT problems \SGCD can
solve more than \ProverN, which, like \SGCD, creates CD proofs. However, \SGCD
fails on five problems that \ProverN can solve. It solves no new problems
(i.e., problems rated with 1.00), but 93\% of the problems that are solvable
at all by a first-order prover and 95\% of the problems that can be solved by
the \name{E} prover.

\subsection{\SGCD in Purely  Goal-Driven Mode and Clausal Tableaux}
\label{sec-exp-tabx}

\SGCD stems from goal-driven provers that can be characterized in terms of the
connection method, model elimination or clausal tableaux and operate in
essence by enumerating proof structures. As Table~\ref{tab-goaldriven}
verifies, in purely goal-driven configurations \SGCD is indeed very similar to
these provers and can roughly solve the same problems as several of them taken
together.

\begin{table}[h]
  \centering
  \caption{The performance of \SGCD in purely goal-driven mode.}
  \vspace{4pt}
  \label{tab-goaldriven}             
  \begin{tabular}{l@{\hspace{1.0em}}r}
    \name{Corpus \TPTPCDT} & 196\\
    \name{SGCD-1+2+3+7} & 176\\\midrule   
    \name{SGCD-G1} & 81\\
    \name{SGCD-G2} & 65\\
    \name{SGCD-G1+G2} & 89\\\midrule
    \name{TABX} & 92\\
    \name{SGCD-G1+G2} $\setminus$ \name{TABX} & 3 \\
    \name{TABX} $\setminus$ \name{SGCD-G1+G2} & 6 \\\midrule
    \CMProver & 89\\
    \name{SETHEO 3.3} & 65\\
    \name{S-SETHEO} & 66\\
    \name{lazyCoP 0.1} & 42\\
    \name{SATCoP 0.1} & 59\\
  \end{tabular}
\end{table}

The table shows, like Table~\ref{tab-solnum} the cardinalities of subsets of
problems in corpus \TPTPCDT. The value for \name{SGCD-1+2+3+7}, the set of
problems solvable by \SGCD in at least one of four configurations that blend
goal- and axiom-driven operation, is transferred from Table~\ref{tab-solnum}.
Rows \name{SGCD-G1} and \name{-G2} show the number of solved problems for two
configurations of \SGCD that are purely goal-driven with generators over tree
size and height, respectively.\footnote{With hardware and timeout settings as
  described above in footnote~\ref{foot-exp-cond}.} Row \name{SGCD-G1+G2}
shows the number of problems from that was solved by at least one of these
configurations. The set \name{SGCD-1+2+3+7} includes all of these.

Row~\name{TABX} shows for comparison the number of
problems that was solved by at least one of the following provers: \CMProver
\cite{cw:pie:2016,cw:pie:2020} in at least one of several considered
configurations, \name{SETHEO~3.3} \cite{setheo:3.3:1997}, \name{S-SETHEO}
\cite{s-setheo:2001}, \name{lazyCoP~0.1} \cite{lazycop:2021} and
\name{SATCoP~0.1} \cite{satcop:2021}.\footnote{For \CMProver according to
  \url{http://cs.christophwernhard.com/pie/cmprover/evaluation_201803/tptp_neq.html},
  for the other systems according to the \name{ProblemAndSolutionStatistics}
  document of \name{TPTP 8.0.0}.} The next two rows indicate the numbers of
problems that can be solved either by \SGCD in at least one of the two
goal-driven configurations or by at least one of the five contributors to
\name{TABX}, but not both. The remaining rows show the number of problems
solved individually by the contributors to \name{TABX}.\footnote{Also
  \name{leanCoP~2.1} \cite{leancop} was tried. With a single exception, all
  \TPTPCDT problems are provided in the \TPTP CNF format, which is not
  accepted by \name{leanCoP}. Experiments in the same settings as \SGCD on
  conversions to \TPTP FOF format (with axioms and goal conjecture
  distinguished) lead to 50~solved problems, subsumed by those of
  \name{TABX}.}

In summary, from the viewpoint of the goal-driven provers based on clausal
tableaux or the conventional connection method, \SGCD in purely goal-driven
configurations is very similar to a strong prover of that family and the
possibility to blend goal- and axiom-driven proceeding in \SGCD is a
substantial improvement, roughly doubling the number of \TPTPCDT problems that
can be solved.

\subsection{68 Theses by \Lukasiewicz in a Single Axiom-Driven \SGCD Run}
68 of the \TPTPCDT problems are from the same source: Theses that follow from
an axiomatization of propositional logic with three axioms by \Lukasiewicz and
are proven in his book \cite{luk-book}. They were used extensively with \OTTER
\cite{otter} by Larry Wos
\cite{wos:bledsoe:91,mccune:wos:cd:1992,wos:resonance:95,wos:combining:96}. In
a purely axiom-driven configuration \SGCD finds these 68~theses all together
in about 2.5 min. As parameters that relate to the theses, the configuration
only considers maximal term dimensions among all theses. The generator is over
tree size and the cache size is limited to 3000 solutions, trimmed according
to dimensions of the MGTs. The procedure exhausts after level (i.e., tree
size) 240. All theses were found at levels below 150, 56 of them among the
3000 cached solutions and the remaining 12 ones among 126,839 residual deleted
solutions. \OTTER solved the problems in a single run after the introduction
of weight templates \cite{wos:bledsoe:91,wos:resonance:95}. The focus in
\cite{wos:combining:96} was to find short overall proofs for subsets of the
theses that represent axiom systems. As shown in Table~\ref{tab-68}, our
initial results can not compete with the carefully developed proofs from
\cite{wos:combining:96}, except with respect to tree size.

\begin{table}[h]
  \centering
\caption{Dimensions of the combined proofs of axiom systems (after
  \cite{wos:combining:96}). In \cite{wos:combining:96} compacted size is
  termed \name{length} and height \name{level}.}
\label{tab-68}
  \setlength{\tabcolsep}{4pt}
\begin{tabular}{lrr}
  Axiom System  & Proof by \SGCD & Proof from \cite{wos:combining:96}\\\midrule
\name{Church} &  $\ddim{43}{125}{19}$ & $\ddim{21}{218}{15}$ \\
\name{Frege} &  $\ddim{62}{277}{19}$ & $\ddim{28}{473}{17}$\\
\name{Hilbert}  & $\ddim{44}{196}{17}$ & $\ddim{23}{483}{15}$\\
\name{Alt. \Lukasiewicz} & $\ddim{48}{138}{17}$ & $\ddim{24}{277}{15}$\\
\name{Wos} & $\ddim{57}{170}{18}$ & $\ddim{25}{335}{16}$\\
\end{tabular}
\end{table}

\subsection{Dimensions of Proofs Returned by \SGCD}
\label{sec-exp-dim}

As already mentioned, the proofs for CD problems by \ProverN in its default
configuration can be easily translated to D-terms, allowing direct size
comparisons. Table~\ref{tab-sizes} shows for the 163 \TPTPCDT problems
solvable by both \name{SGCD-1+2+3+7} (as specified in
Sect.~\ref{sec-exp-sgcd}) and \ProverN the compacted size, tree size and
height of the proofs by \SGCD and \ProverN, aggregated as average and median
values. For \SGCD, for each problem and size measure the minimal value of all
the up to four proofs obtained with the four configurations of
\name{SGCD-1+2+3+7} is considered. The $^n$-superscript indicates values after
n-simplification \cite{cwwb:lukas:2021}, which has a strong effect on the
proofs by \ProverN but only a very minor effect on those by \SGCD. In summary,
the table indicates that \SGCD returns smaller proofs than \ProverN in its
default mode, moderately smaller with respect to compacted size and height,
and drastically smaller with respect to tree size.
\begin{table}[h]
  \centering
  \caption{Size of CD proofs obtained by \SGCD and \ProverN.}
  \label{tab-sizes}
\setlength{\tabcolsep}{4pt}
  \begin{tabular}{lrrrrrrrr}
& C-avg & C-med & T-avg & T-med & H-avg & H-med\\\midrule
\name{SGCD-1+2+3+7}$^n$ & 15.80 & 12 & 29.36 & 17 & 8.52 & 6\\
\name{SGCD-1+2+3+7}     & 15.83 & 13 & 29.36 & 17 & 8.52 & 6\\
\name{Prover9}$^n$    & 23.09 & 18 & 19501.99 & 40 & 13.69 & 12\\
\ProverN              & 28.37 & 21 & 194736.83 & 93 & 16.90 & 13\\

  \end{tabular}
\end{table}

\subsection{PSP Structure Enumeration and Proofs with Small Compacted Size}
\label{sec-exp-psp}

We consider the objective to find proofs with small compacted size. \CCS can
be applied in a configuration that returns proofs with ascertained minimal
compacted size by exhaustive search \cite{cw:ccs}. This succeeds for about
44\% problems of the \TPTPCDT corpus. The remaining problems have to be
tackled with techniques that may not yield proofs of ascertained minimal
compacted size, but typically proofs with small compacted size. Aside of the
configurations of \SGCD based on enumeration by tree size or height discussed
in the previous sections, there is another prospective candidate, a novel
proof enumeration strategy for \SGCD called \name{PSP} (indicating
\name{Proof-SubProof}), specified as follows, where \name{level} is understood
in the sense of Sect.~\ref{sec-sgcd}.
\begin{enumerate}[(i)]
\item Structures at level $0$ are the axiom identifiers.
\item Structures at level $n+1$ are all structures $\D(d_1,d_2)$ and
  $\D(d_2,d_1)$ where $d_1$ is a structure at level $n$ and $d_2$ is a (not
  necessarily strict) subterm of $d_1$ or an axiom identifier.
\end{enumerate}
PSP was motivated by an analysis \cite{cwwb:lukas:2021} of Meredith's variant
\cite{meredith:notes:1963} of \Lukasiewicz's completeness proof for his
shortest single axiom for implicational logic \cite{luk:1948}, where it was
observed that most steps in Meredith's proof can be ascribed the relationship
between levels that underlies PSP. The experiments summarized in
Table~\ref{tab-psp} evaluate the potential of \SGCD in configurations with PSP
and with other structure generation methods for finding proofs, with emphasis
on small compacted size.

\begin{table}[h]
\caption{Proofs with small compacted size by \SGCD and \CCS.}
\label{tab-psp}

\begin{tabular}{l@{\hspace{2.0em}}r@{\hspace{1.0em}}r@{\hspace{1.0em}}r}
                         &       & $>$0.00 & $>$0.25\\\midrule
  \name{Corpus \TPTPCDT} &  196  &          45 &           20\\\midrule
  \name{\CCS-MinC}    &   86  & 3 & 0\\
  \name{\SGCD-PSP}       & 153 & 14 & 2\\
  \name{\SGCD-NonPSP}     & 176 & 25 & 5\\\midrule
  \name{\CCS-MinC} $\cap$ \name{\SGCD-PSP}   & 85 & 3 & 0\\
  \name{\CCS-MinC} $\cap$ \name{\SGCD-NonPSP} & 86 & 3 & 0\\
  \name{\CCS-MinC} $\cap$ (\name{\SGCD-PSP = \CCS-MinC})  & 60 & 1 & 0\\
  \name{\CCS-MinC} $\cap$ (\name{\SGCD-NonPSP = \CCS-MinC})& 59 & 1 & 0\\\midrule
  \name{$\lnot$ \CCS-MinC} $\cap$ \name{\SGCD-PSP}   & 68 & 11 & 2\\
  \name{$\lnot$ \CCS-MinC} $\cap$ \name{\SGCD-NonPSP} & 90 & 22 & 5\\
  \name{$\lnot$ \CCS-MinC} $\cap$ (\name{\SGCD-PSP} < \name{\SGCD-NonPSP}) & 44 & 5  & 1\\
  \name{$\lnot$ \CCS-MinC} $\cap$ (\name{\SGCD-NonPSP} < \name{\SGCD-PSP}) & 43 & 17 & 4\\
  \name{$\lnot$ \CCS-MinC} $\cap$ (\name{\SGCD-PSP} = \name{\SGCD-NonPSP}) & 3 & 0 & 0\\
\end{tabular}
\end{table}

\SGCD was there considered in five configurations with PSP structure
generation and in eight ``NonPSP'' configurations, that is, configurations
with structure generation over tree size or height, which included those of
\name{SGCD-1+2+3+7} considered in Sect.~\ref{sec-exp-sgcd}. Conditions were as
described in footnote~\ref{foot-exp-cond}.
Each row in Table~\ref{tab-psp} shows for a set of problems from \TPTPCDT its
cardinality, as well as the number of members with rating $>$ 0.00 and the
number of members with rating $>$ 0.25.
\name{\CCS-MinC} is the set of those problems for which a proof with
ascertained compacted size could be found with \CCS \cite[Sect.~3.1]{cw:ccs}.
\name{\SGCD-PSP} is the set of problems for which a proof could be found by
\SGCD in a PSP configuration, \name{\SGCD-NonPSP} is the set of problems for
which a proof could be found by \SGCD in a NonPSP configuration.
The four rows prefixed by ``\name{\CCS-MinC}~$\cap$'' relate \name{\SGCD-PSP}
and \name{\SGCD-NonPSP} to \name{\CCS-MinC}. The intersection symbol is to be
read as set intersection. (\name{\SGCD-PSP} = \name{\CCS-MinC}) denotes the
subset of problems from \name{\SGCD-PSP} for which a PSP configuration
returned a proof with the ascertained minimal compacted size. The
(\name{\SGCD-NonPSP} = \name{\CCS-MinC}) is the analog for the NonPSP
configurations.

The five rows prefixed by ``$\lnot$ \name{\CCS-MinC}~$\cap$'' relate
\name{\SGCD-PSP} and \name{\SGCD-NonPSP} to the set of problems for which the
minimal compacted size could not be determined by exhaustive search with \CCS,
which are the problems that actually benefit from the enumeration methods and
heuristic restrictions offered by \SGCD. The five rows indicate the
contribution of \SGCD in PSP and NonPSP configurations to increase the number
of solvable problems, taking into account the compacted size of proofs. $\lnot
\name{\CCS-MinC}$ denotes the complement of \name{\CCS-MinC} with respect to
the corpus \TPTPCDT. (\name{\SGCD-PSP} < \name{\SGCD-NonPSP}) denotes the
subset of the problems of \name{\SGCD-PSP} that could either not be proven in
the NonPSP configurations or could be proven in a NonPSP configuration only
with a compacted size that is larger than the smallest compacted size obtained
with a PSP-based configuration. (\name{\SGCD-NonPSP} < \name{\SGCD-PSP}) is
analogous, with the roles of \name{\SGCD-PSP} and \name{\SGCD-NonPSP}
switched. (\name{\SGCD-PSP} = \name{\SGCD-NonPSP}) denotes the set of problems
in both \name{\SGCD-PSP} and \name{\SGCD-NonPSP} such that the smallest
compacted size of proofs by the PSP configurations and the smallest compacted
size of proofs by the NonPSP configurations are equal.

In summary, the table shows that the problems for which the required minimal
compacted proof size can be determined successfully by exhaustive search can
(with one exception) also be proven by \SGCD in both PSP and NonPSP
configurations. The ascertained minimal size is achieved there with \SGCD for
about 70\% of the problems in either family of configurations.
For problems where determining the minimal compacted proof size by exhaustive
search with \CCS failed, configurations of both families succeed in many
cases, roughly doubling the number of proven problems compared to those for
which the required minimal compacted size can be determined. PSP
configurations succeed there for a subset of about 3/4 of the problems on
which the NonPSP configurations succeed. When the compacted size of the proofs
is taken into account, the contributions of both configuration families are of
the same order, each family yielding for about one half of the solvable
problems a proof with smaller compacted size than the smallest obtained with
the other family.

For PSP, the experiments show that -- although motivated by observations on a
proof by a human for a particular problem -- it is applicable as a general
structure generation technique to prove a large portion of problems from
corpus \TPTPCDT and, in particular, is useful to complement other structure
generation techniques when the objective is to find proofs with small
compacted size.

\subsection{A Short Proof for \Lukasiewicz's Shortest Axiom}
\label{sec-exp-luk}

The completeness of \Lukasiewicz's shortest single axiom for the implicational
propositional calculus has been proven originally \cite{luk:1948} with
dimensions $\ddim{34}{585}{29}$ and in \cite{meredith:notes:1963} refined to
$\ddim{33}{669}{29}$ \cite{cwwb:lukas:2021}. Representations of these historic
formal proofs by logicians are included in \CDTOOLS. Guided by an analysis of
them, slightly shorter proofs were obtained in \cite{cwwb:lukas:2021}. With
\SGCD, the much shorter proof shown in Fig.~\ref{fig-luk-short} was found,
whose dimensions are $\ddim{29}{92}{22}$. It was obtained by some interaction
with the \CDTOOLS system as follows: First, \SGCD in a PSP-based configuration
returned a relatively short proof of the most difficult of the three involved
subproblems, \tp{LCL038-1}, with dimensions $\ddim{22}{64}{22}$, which compare
to $\ddim{31}{491}{29}$ in \cite{meredith:notes:1963}. This proof was then
supplemented with proofs of the other two subproblems, which were found
through enumerating by tree size and selecting according to the compacted size
of all three proofs taken together. The structure of the PSP-based proof of
\tp{LCL038-1} with dimensions~$\ddim{22}{64}{22}$, step~$7$ in
Fig.~\ref{fig-luk-short}, is shown as DAG in Fig.~\ref{fig-luk-short-graph}.

\begin{figure}[h]
  \centering
  \caption{$\ddim{29}{92}{22}$-proof of the completeness of \Lukasiewicz's
    axiom in the Meredith's notation \cite{meredith:notes:1963}.}
  \label{fig-luk-short}
  \small
  \vspace{4pt}
  \begin{tabular}{rl}
1. & $\g{CCCpqrCCrpCsp}$\\
2. & $\g{CCCCpqCrqCqsCtCqs} \mereq \f{D}11$\\
3. & $\g{CCCpCqrCCsqCtqCuCCsqCtq} \mereq \f{D}12$\\
4. & $\g{CCCpCqrCstCCqtCst} \mereq \f{D}\f{D}\f{D}\f{D}1\f{D}1\f{D}1\f{D}1\f{D}\f{D}\f{D}\f{D}131\mathrm{n}11\mathrm{n}1$\\
5. & $\g{CCCCpqCrqCCCsCptCrquCvCCCsCptCrqu} \mereq \f{D}1\f{D}\f{D}414$\\
6. & $\g{CCCpqpCrp} \mereq \f{D}\f{D}31\mathrm{n}$\\
*7. & $\g{CCpqCCqrCpr} \mereq \f{D}\f{D}\f{D}\f{D}1\f{D}\f{D}55\mathrm{n}1\mathrm{n}1$\\
*8. & $\g{CCCpqpp} \mereq \f{D}\f{D}\f{D}426\mathrm{n}$\\
*9. & $\g{CpCqp} \mereq \f{D}\f{D}26\mathrm{n}$\\
  \end{tabular}
\end{figure}

\begin{figure}[p]
  \centering
  \caption{The structure of the proof of step~7 from Fig.~\ref{fig-luk-short}
    (\tp{LCL038-1}) as DAG. Node numbers indicate the correspondence to that
    figure. A dashed arrow indicates that the actual formula used as minor
    premise plays no role to determine the conclusion, which is indicated by
    ``$\n$'' in Meredith's notation. (See also the discussion of
    n-simplification in \cite{cwwb:lukas:2021}.)}
  \label{fig-luk-short-graph}
  \vspace{-10pt}
  {\normalfont \scalebox{0.345}{\input{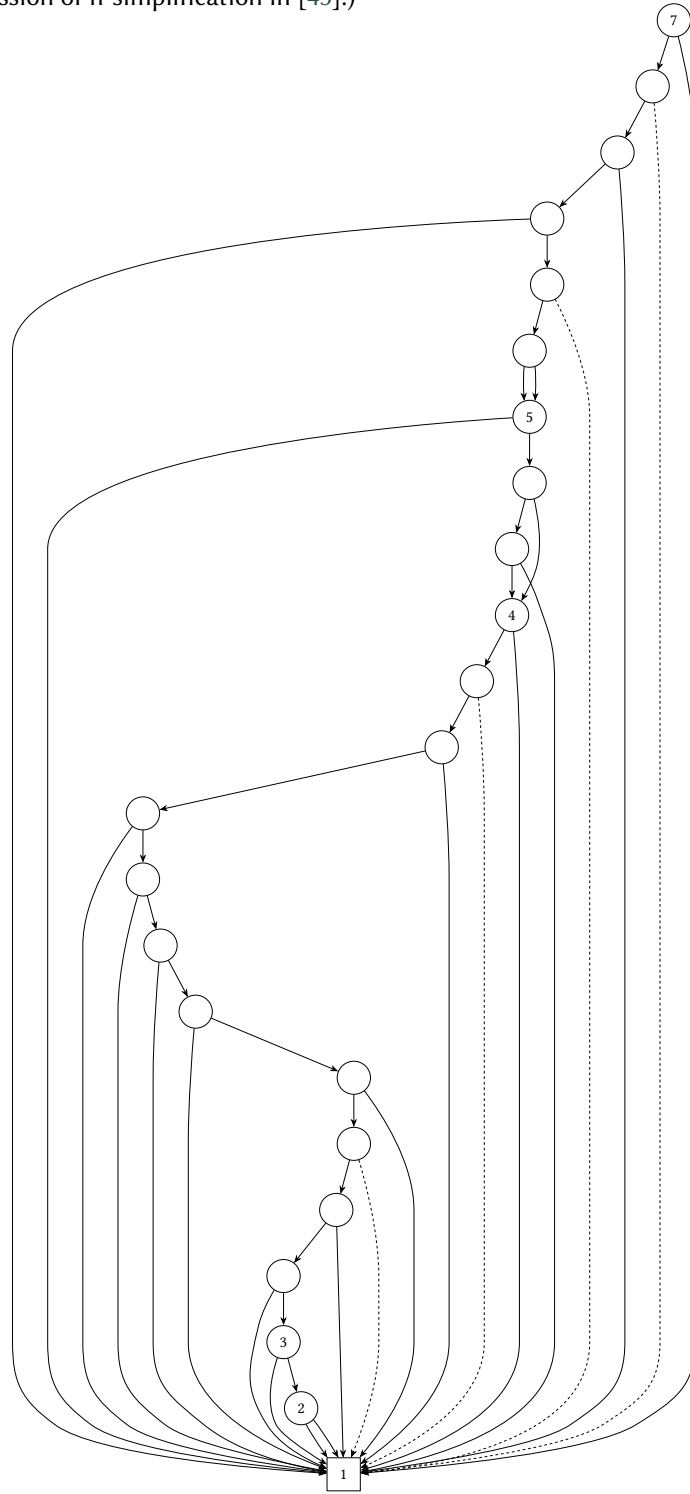}}}
\end{figure}

\subsection{Proof Structure Compression by Tree Grammars and by Combinators}
\label{sec-exp-compress}

Meredith's notation used in Fig.~\ref{fig-luk-short} can be read as a tree
grammar where the line numbers except of 1 for the axiom are nonterminals. For
each nonterminal the grammar generates exactly one D-term. The grammar
represents the proof as a DAG, where the number of written $\f{D}$s is the
compacted size. Stronger compressions can be achieved by grammars where the
nonterminals are permitted to contain parameters \cite{lohrey:survey:2015},
which has been considered before in proof theory for term substitutions in
formulas \cite{hetzl:tree:2012}. \SGCD supports experimenting with such
compressions applied to the proof structure through an interface to
\name{TreeRePair} \cite{lohrey:treerepair:2013}, an advanced tree compression
system originally targeted at XML document trees. The right column of
Fig.~\ref{fig-grammar} shows a grammar obtained with this system for the
subproof of $7$ from Fig.~\ref{fig-luk-short}.

\begin{figure}[h]
\caption{A tree grammar compression of the subproof of step~7 from
  Fig.~\ref{fig-luk-short} (\tp{LCL038-1}) (with $\n$ replaced by $1$)
  obtained by \name{TreeRePair} invoked via \CDTOOLS.}
\label{fig-grammar}
  \centering
  $\begin{array}{rllcl}
  1. & \g{CCCpqrCCrpCsp}\\
  2. & \g{CCCCpqrCCrpCspt} \imp \g{t} & A_2(v) & \rightarrow & \D(v,1)\\
  3. & \g{CCpqr} \imp \g{CCrpCsp} & A_3(v) & \rightarrow & \D(1,v)\\
  4. & \g{CCpqr} \imp \g{CCCsprCtr} & A_4(v) & \rightarrow & A_3(A_3(v))\\
  5. & \multicolumn{4}{l}{\g{CCCCpqrCCrpCspCCCCtuvCCvtCwtCCCCxyzCCzxCaxb} \imp \g{b}}\\
  && A_5(v) & \rightarrow & A_2(A_2(A_2(v)))\\
  6. & \g{CCCpCqrCstCCqtCst} & A_6 & \rightarrow & A_5(A_4(A_4(A_5(A_4(A_2(1))))))\\
  7. & \g{CCCCpqCrqCCCsCptCrquCvCCCsCptCrqu}\;\; & A_7 & \rightarrow & A_3(\D(A_2(A_6),A_6))\\
  *8. & \g{CCpqCCqrCpr} & \mathit{Start} & \rightarrow & A_5(A_3(A_2(\D(A_7,A_7))))
\end{array}$
\end{figure}

The left column of Fig.~\ref{fig-grammar} shows lemmas that correspond to the
respective nonterminals. For a parameter-free nonterminal it is the MGT of the
generated D-term, for a nonterminal with parameters it is a Horn clause. To
represent these Horn clauses we extend the Polish notation for the lemmas
corresponding to the proof steps. For example, $\g{CCpqr} \imp \g{CCrpCsp}$ of
step~3 reads in our first-order representation as $\P((p\fimp q)\fimp r) \imp
\P((r\fimp p)\fimp (s\fimp p))$. A common size measure for such grammars is
the sum of the number of edges on the right-hand sides of the productions,
24~in the example. It compares to the doubled value of the compacted size,
$22*2 = 44$, because each inner node of the DAG underlying the compacted size
has two outgoing edges.

\begin{figure}
  \centering
  \caption{A combinatory compression of the structure of the proof of step~7
    from Fig.~\ref{fig-luk-short} (\tp{LCL038-1}) as DAG, obtained via the
    grammar compression shown in Fig.~\ref{fig-grammar}.}
  \label{fig-compress-luk}
  \vspace{2pt}
  {\normalfont \scalebox{0.345}{\input{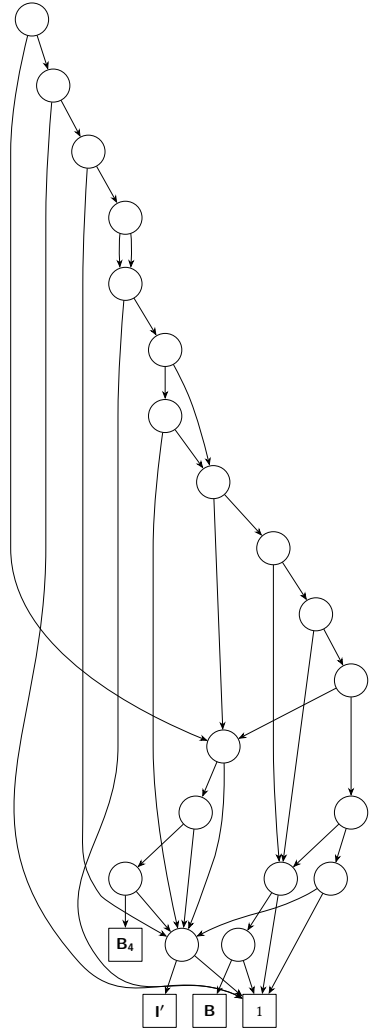}}}
\end{figure}

\CDTOOLS also supports a novel way of proof structure compression introduced
in \cite{cw:ccs}, which works by permitting combinators in the D-terms. It can
be used as basis for proof search over compressed structures and it lets
uncompressed proof structures appear as normal forms, in the technical sense
of combinator reduction, of compressed proof structures. Tree grammar
compressions can be converted to combinator compressions with techniques from
functional programming languages \cite{peytonjones:87}, which is described in
\cite{cw:ccs} and implemented in \CDTOOLS. Figure~\ref{fig-compress-luk} shows
an example, the combinatory compression obtained from the grammar compression
shown in Fig.~\ref{fig-grammar}. Its dimensions are $\ddim{19}{119}{15}$,
which compare to $\ddim{22}{64}{22}$, the dimensions of the expanded structure
shown in Fig.~\ref{fig-luk-short-graph}. The involved combinators are
associated with the following rewriting rules.
\[
\begin{array}{lcl}
\D(\D(\c{I'}, x), y) & \rimp & \D(y,x)\\
\D(\D(\D(\c{B},x),y),z) & \rimp & \D(x,\D(y, z))\\
\D(\D(\D(\D(\c{B_4},x),y),z),u) & \rimp &  \D(x,\D(y,D(z,u)))
\end{array}
\]
Rewriting the tree representation of Fig.~\ref{fig-compress-luk} with these
rules yields the tree representation of Fig.~\ref{fig-luk-short-graph}.
Combinatory compression as provided by \CDTOOLS is discussed in more depth in
\cite{cw:ccs}.

\pagebreak
\section{Conclusion}
\label{sec-conclusion}
\enlargethispage{8pt}

A Prolog library for experimenting with CD in ATP has been presented. It puts
an ATP-oriented formal view on CD that is centered around proof structures
represented as terms \cite{cwwb:lukas:2021} into practice. \SGCD, an included
first-order prover is specialized on CD problems and competes for these with
the lower end of general state-of-the-art provers. It is based on a simple but
so far not fully explored paradigm, enumerating proof structures in a way that
combines two modes: a goal-driven mode, as known from systems based on the
connection method or clausal tableaux, and an axiom-driven mode that generates
lemmas in a level-by-level way. In generic configurations it produces
comparatively short proofs. With special configurations very short proofs of
difficult problems can be found, for which the systematic detection of
shortest proofs seems too hard.

We conclude the presentation with some remarks on related work and
perspectives.
Similarly to \SGCD, an implementation \cite{cw:ekrhyper:2007} of hypertableaux
\cite{hypertableaux} creates lemmas axiom-driven level-by-level, however much
less flexibly configurable and incorporating goal-driven phases only in a
rudimentary way, by immediate termination before a level is completed if the
goal is generated.
A recent enhancement of \E by machine learning supplements a classification
based on \emph{derivation history} as clause selection guidance
\cite{suda:history:2021}. This may be viewed as introducing into \E a bit of
the structure generating approach, where the derivation history in form of
D-terms is the main guidance.
That the approach of \SGCD yields particularly small proofs should be of
general relevance for applications that are based on proof structures emitted
by reasoners, for example to produce ``good'' explanations
\cite{alrabbaa:goodproofs:2021} and for interpolation
\cite{cw:interpolation:2021}.
In extreme configurations, \SGCD acts purely goal-driven or purely
axiom-driven. Purely goal-driven configurations perform very similarly to
systems based on the connection method or clausal tableaux, which, however,
are for the CD problems much weaker than state-or-the-art provers. Purely
axiom-driven configurations may be useful as a basis for finding
mathematically interesting theorems from axioms.
\CDTOOLS is a versatile platform for experimental studies of various kinds
such as, for example, investigating grammar- and combinator-based compressions
of proof structures, which are both stronger than just compressing trees into
DAGs. Combinator-based compression is not only applicable to shorten a given
proof, but also for practical proof search, realizing features known from the
connection structure calculus \cite{eder:cs:1989,bibel:eder:1993} but not
implemented previously. \CDTOOLS made the initial elaboration and evaluation
of this observation possible \cite{cw:ccs}. This involved writing a second
structure generating prover, \CCS, which is now also included in the system.

\begin{acknowledgments}
  The author thanks Wolfgang Bibel for inspiration, and him as well as
  anonymous reviewers of a previous version for helpful suggestions to improve
  the presentation. Funded by the Deutsche Forschungsgemeinschaft (DFG, German
  Research Foundation) -- Project-ID~457292495. The work was supported by the
  North-German Supercomputing Alliance (HLRN).
\end{acknowledgments}

\bibliography{biblukas03ed}

\end{document}